\documentclass[a4paper,fleqn,usenatbib]{mnras}

\usepackage[T1]{fontenc}
\usepackage{ae,aecompl}

\usepackage[normalem]{ulem}
\usepackage{amsmath}
\usepackage{graphicx} 
\usepackage{lscape}
\usepackage{indentfirst}
\usepackage{enumitem}
\usepackage{xspace}

\usepackage{amssymb}
\usepackage[dvipsnames]{xcolor}
\usepackage{url}
\usepackage[flushleft]{threeparttable}

\newcommand {\bc}{\begin {center}}
\newcommand {\ec}{\end {center}}
\newcommand {\be}{\begin {equation}}
\newcommand {\ee}{\end {equation}}
\newcommand {\beq}{\begin {eqnarray}}
\newcommand {\eeq}{\end {eqnarray}}
\newcommand {\comment}[1]{}

\renewcommand{\d}{{\rm d}}

\newcommand {\ergs}{{\rm erg\ \rm s^{-1}}}

\title[Pulse profiles fluctuations]
{
Flickering pulsations in bright X-ray pulsars:\\
the evidence of gravitationally lensed and eclipsed accretion column
}
\author[A.~Mushtukov et al.] 
{
Alexander A. Mushtukov$^{1}$\thanks{E-mail: alexander.mushtukov@physics.ox.ac.uk (AAM)},
Albert Weng$^{2}$\thanks{E-mail: albertweng1213@gmail.com (AW)},
Sergey S. Tsygankov$^{3}$,
Ilya A. Mereminskiy$^{4}$
\\ 
$^1$ Astrophysics, Department of Physics, University of Oxford, Denys Wilkinson Building, Keble Road, Oxford OX1 3RH, UK\\
$^2$ Fairmont Preparatory Academy, 2200 W Sequoia Ave, Anaheim, CA 92801, USA \\ 
$^3$ Department of Physics and Astronomy,  FI-20014 University of Turku, Finland \\
$^4$ Space Research Institute, Russian Academy of Sciences, Profsoyuznaya 84/32, 117997 Moscow, Russia\\
} 

\pubyear{2023}

\begin{document}
\label{firstpage}
\pagerange{\pageref{firstpage}--\pageref{lastpage}}
\maketitle


\begin{abstract} 
It is expected that extreme mass accretion rate onto strongly magnetised neutron star results in appearance of accretion columns above stellar surface.
For a distant observer, rotation of a star results in periodic variations of X-ray flux. 
Because the mass accretion rate fluctuates around the average value, the pulse profiles are not stable and demonstrate fluctuations as well.
In the case of bright X-ray pulsars, however, pulse fluctuations are not solely attributed to variations in the mass accretion rate. 
They are also influenced by the variable height of the columns, which is dependent on the mass accretion rate.
This study delves into the process of pulse profile formation in bright X-ray pulsars, taking into account stochastic fluctuations in the mass accretion rate, the corresponding variations in accretion column geometry and gravitational bending.
Our analysis reveals that potential eclipses of accretion columns by a neutron star during their spin period should manifest specific features in pulse profile variability. 
Applying a novel pulse profile analysis technique, we successfully detect these features in the bright X-ray transient V~0332+53 at luminosities $\gtrsim 2\times 10^{38}\,\ergs$. 
This detection serves as compelling evidence for the eclipse of an accretion column by a neutron star.
Detection of the eclipse places constraints on the relation between neutron star mass, radius and accretion column height.
Specifically, we can establish an upper limit on the accretion column height, which is crucial for refining theoretical models of extreme accretion.
\end{abstract}

\begin{keywords}
accretion -- accretion discs -- X-rays: binaries -- stars: neutron -- stars: oscillations
\end{keywords}

\section{Introduction}
\label{sec:Intro}

Accreting strongly magnetised neutron stars (NSs) manifest themselves as X-ray pulsars (XRPs, see \citealt{2022arXiv220414185M} for review).
Magnetic field ($B$-field) at the NS surface in XRPs is typically $\gtrsim 10^{12}\,{\rm G}$.
Strong field of a NS disturbs accretion flow at distance $\sim 10^7 - 10^8\,{\rm cm}$, which is called the magnetospheric radius, and  shapes the geometry of accretion flow within it directing material towards small regions located near the magnetic poles of a NS. 
Accretion flow is inevitably decelerated at the NS surface or in close proximity to the surface.
The geometry of the emitting region depends on the mass accretion rate.
In the case of relatively low mass accretion rates  (sub-critical regime, $\lesssim 10^{17}\,{\rm g\,s^{-1}}$), the flow reaches the stellar surface and is decelerated in the atmosphere of a NS \citep{1969SvA....13..175Z}, which results in hot spot geometry of the emitting regions. 
At high mass accretion rates (super-critical regime, $\gtrsim 10^{17}\,{\rm g\,s^{-1}}$), the dynamics of accretion flow is affected by the radiation pressure and the flow is decelerated in radiation dominated shock above NS surface \citep{1976MNRAS.175..395B,1981A&A....93..255W,2015MNRAS.454.2539M,2023MNRAS.524.4148A}.
Shock wave above the surface and the sinking region below it form an accretion column - structure supported by the radiation pressure and confined by extremely strong magnetic field.
The accurate numerical model of accretion columns is still developing and requires accounting for many aspects, including magneto-hydrodynamics \citep{2023MNRAS.524.2431S}, radiative transfer in a strong magnetic field (see \citealt{1992herm.book.....M} for review, and \citealt{2022MNRAS.517.4022S} for the recent results) and possibly energy losses due to creation of electron-positron pairs and further neutrino emission \citep{2018MNRAS.476.2867M,2019MNRAS.485L.131M,2023MNRAS.522.3405A}. 

\begin{figure}
\centering 
\includegraphics[width=5.3cm]{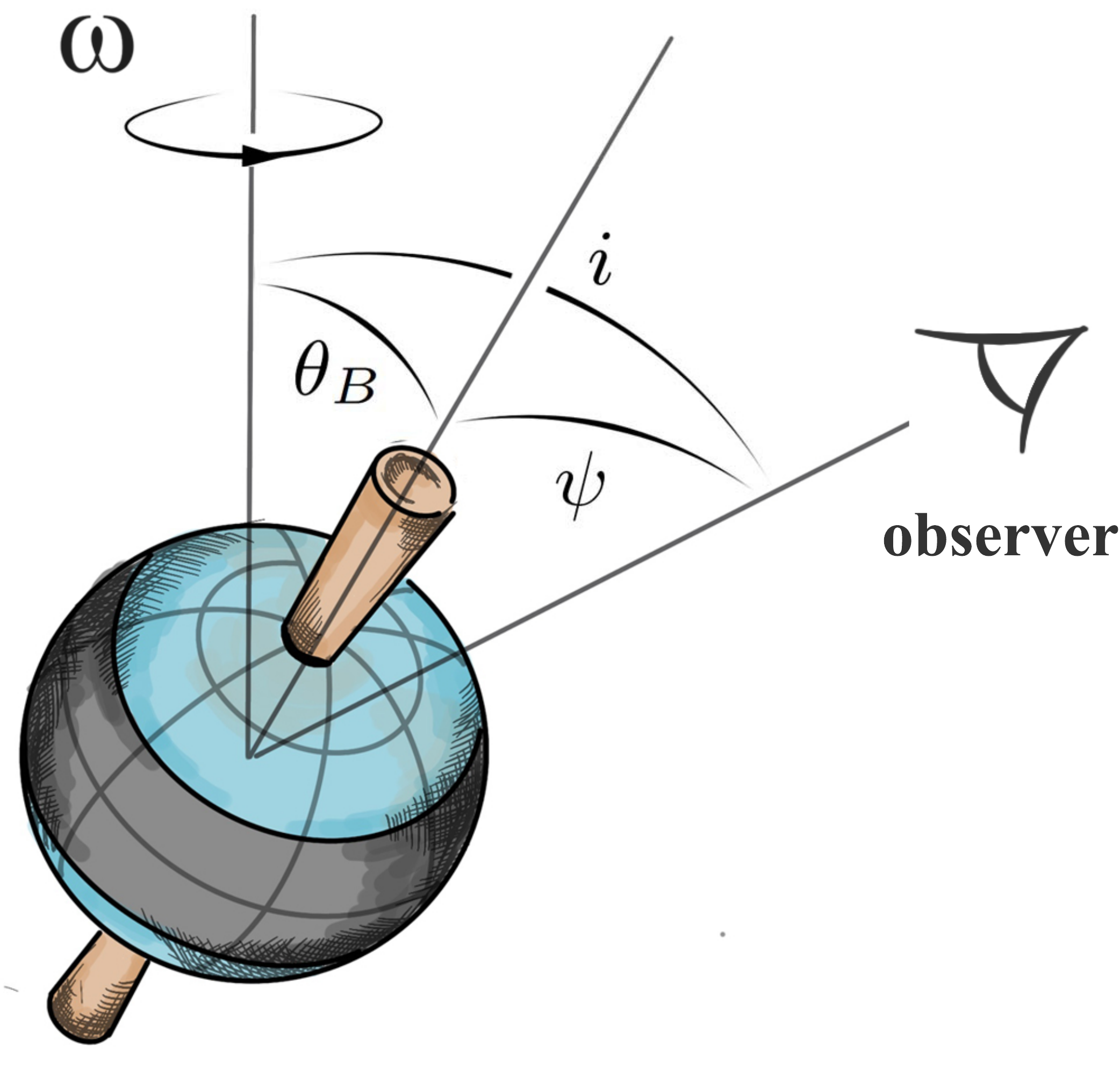} 
\caption{
    Schematic illustration of the emission region geometry {in super-critical XRP} considered in the paper. 
    {X-ray photons are initially produced by extended accretion column. 
    A fraction of X-ray radiation is intercepted by a NS and reflected/reprocessed by the atmosphere.
    The flux detected by a distant observer is composed of direct flux from the columns and the flux reflected by the atmosphere.
    Both direct and reflected components are subjects of gravitational bending. 
    Rotation geometry in the observers reference frame is determined by the inclination of a NS $i$ and the magnetic obliquity $\theta_B$.}
}
\label{pic:scheme_rot}
\end{figure}

The formation mechanism of the beam pattern is anticipated to differ between sub-critical and super-critical XRPs \citep{1973A&A....25..233G}. In the sub-critical regime, a majority of X-ray photons originate from hot spots, whose geometry exhibits only marginal dependence on the mass accretion rate. Conversely, in the super-critical regime, photons are emitted by the walls of the accretion column, where the column height correlates with the mass accretion rate \citep{1976MNRAS.175..395B} and is contingent on the strength of the neutron star's magnetic field \citep{2015MNRAS.454.2539M}. 
A fraction of the X-ray radiation emitted by the columns is intercepted and reprocessed by the atmosphere of a NS \citep{2013ApJ...777..115P,2015MNRAS.452.1601P}.
Thus, in the case of super-critical accretion, a distant observer detects a combination of direct flux from the accretion column and flux reflected by the neutron star (see Fig.\,\ref{pic:scheme_rot}). 
Both components of the pattern undergo gravitational light bending in the vicinity of a NS \citep{1988ApJ...325..207R,2001ApJ...563..289K,2018MNRAS.474.5425M,2023MNRAS.tmp.3130M}.

Mass accretion rate in all X-ray binaries, including XPRs, is known to be fluctuating on different time scales. 
In particular, variations of the mass accretion rate can be generated in accretion disc due to the stochastic nature of viscosity \citep{1997MNRAS.292..679L,2019MNRAS.486.4061M}.
The viscosity in the accretion disc can be caused by magnetic dynamo that generates a poloidal magnetic field component in a random fashion \citep{1991ApJ...376..214B,1995ApJ...446..741B}.
The process is stochastic by its nature and results in viscosity fluctuations on time scales close to the local dynamical time scales.
The local viscosity fluctuations result in local fluctuations of the mass accretion rate at the inner disc radius and, further, at the NS surface. 
Fluctuations in the mass accretion rate onto the NS surface give rise to variations in accretion luminosity, observable as fluctuations in X-ray energy flux by a distant observer. 
In the case of super-critical XRPs with accretion columns, fluctuating mass accretion rates induce variations in the emission region geometry — specifically, the accretion column height and the illuminated fraction of a NS surface.
The dynamic changes in the emission region geometry, along with the corresponding fluctuations in the beam pattern, naturally contribute to additional stochastic variability in X-ray energy flux. 
The relative fluctuations in X-ray energy flux depend on the observer's viewing angle, resulting in an anticipated phase-dependent variability in X-ray flux. While phase-dependent relative root mean square (rms) of a pulse profile has been previously observed in super-critical XRPs \citep{2007AstL...33..368T}, it has yet to be fully explained.

In this paper, we investigate how fluctuations of accretion column geometry affects fluctuations of X-ray energy flux and pulse profile stability in super-critical XRPs.
To test out theoretical model, we analyze fluctuations of pulse profile in bright X-ray transient V~0332+53 in super-critical regime of accretion.
Magnetic field at the NS surface in V~0332+53 ($\sim 3\times 10^{12}\,{\rm G}$) is known from detected cyclotron scattering feature at $\sim 30\,{\rm keV}$ and its harmonics \citep{2005ApJ...634L..97P}. 
Variations of cyclotron line centroid energy provide an evidence of accretion column existence at $L>10^{37}\,\ergs$ \citep{2006MNRAS.371...19T} and transition to the super-critical regiome of accretion at $\sim 10^{37}\,\ergs$ \citep{2017MNRAS.466.2143D}.

\begin{figure*}
\centering 
\includegraphics[width=16.cm]{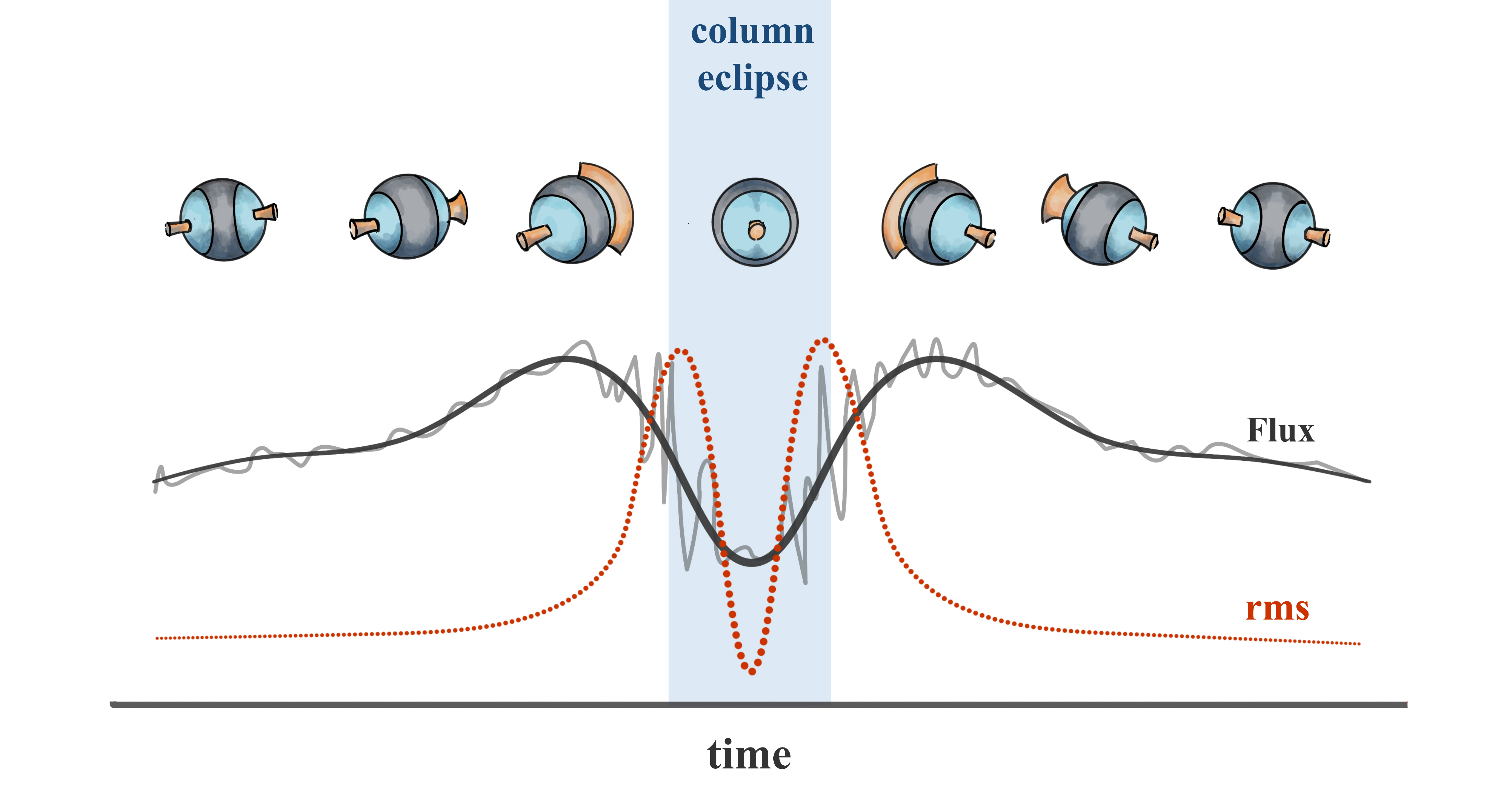} 
\caption{
The anticipated behavior of X-ray energy flux (averaged over many spin periods, denoted by the black line, and the actual flux represented by the grey line) and the phase-resolved rms of X-ray flux (depicted by the red dotted line) during the eclipse of an accretion column by a NS.
During the column eclipse, a drop in X-ray flux is expected. Immediately before and after the eclipse, the direct flux from the accretion column is gravitationally lensed by the NS and may exhibit significant fluctuations, even under conditions of small variations in the mass accretion rate and the corresponding height/luminosity of the accretion column.
}
\label{pic:scheme}
\end{figure*}

\section{Model set up}
\label{sec:Model}

\subsection{Geometry of the emitting region}

\begin{figure}
\centering 
\includegraphics[width=8.5cm]{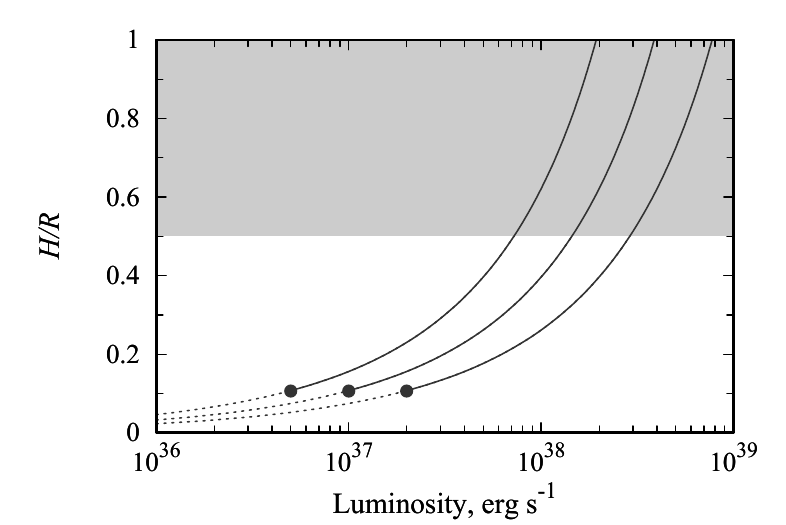} 
\caption{
Approximate dependencies of relative accretion column height on the accretion luminosity calculated according (\ref{eq:L2H_relation}) under the assumption $\kappa_\perp/\kappa_{\rm T}=0.5,\,1,\,2$ (from left to right). 
The black dot shows the expected value of the critical luminosity \citep{2015MNRAS.447.1847M}.
The gray region corresponds to accretion columns heights ruled out by our observational results.
}
\label{pic:AC_height}
\end{figure}

We focus on XRPs in at super-critical regime of accretion ($\dot{M}\gtrsim 10^{17}\,{\rm g\,s^{-1}}$) when X-ray photons are produced by accretion columns.
We assume that a NS has a spherical shape, which is a good approximation for slowly rotating ($P\gtrsim 0.1\,{\rm s}$) NSs.
The columns above the poles of a star are modeled by cylinders of a given height $H\lesssim R$ and radius $r_{\rm c}$.
Accretion column height is assumed to the related to accretion luminosity $L$ and surface magnetic field strength (see, e.g., \citealt{2015MNRAS.454.2539M} and Fig.\,\ref{pic:AC_height}):
\beq\label{eq:L2H_relation}
{L}\sim 2\times 10^{39}\,\left(\frac{\kappa_{\rm T}}{\kappa_\perp}\right)f\left(\frac{H}{R}\right)\,\,\ergs, 
\eeq
where $\kappa_{\rm T}\approx 0.34\,{\rm cm\,g^{-1}}$ is the Thomson electron scattering opacity, $\kappa_\perp$ is the scattering opacity in a strong magnetic field in the direction orthogonal to the field lines, and 
\beq\label{eq:f}
f(x)\equiv \ln(1+x)-\frac{x}{1+x}.
\eeq
The brightness distribution over the column height is described by the function $g(h)$.
In this paper, we assume that the brightness of the accretion column is homogeneously distributed over the height, i.e. $g(h)={\rm constant}$.
The photons leaving the walls of the column are beamed towards the NS surface due to the scattering by free-falling material (see, e.g. \citealt{1976SvA....20..436K,1988SvAL...14..390L,2013ApJ...777..115P}).
The radiation energy from a unit surface area located at the height $h$ into the unit solid angle of the direction $(\alpha,\phi)$ is given by $F_\mathrm{out}f(\alpha,\phi)g(h)$, where $F_\mathrm{out}$ is the 
the flux leaving the accretion column wall.
Following \citealt{2013ApJ...777..115P} we describe the angular distribution of X-ray flux at the walls of accretion columns as
\beq
\label{angle_distr_column}
f(\alpha,\phi)\propto\frac{3D^4}{7\pi\gamma}
\left(1+2D\sin\alpha\cos\phi\right)
\sin^2\alpha\cos\phi
\eeq
where $\beta=v/c$ is the dimensionless velocity in units of the speed of light, 
$\gamma=(1-\beta^2)^{-1/2}$ is the Lorentz factor, 
and $D=[\gamma(1-\beta\cos{\alpha})]^{-1}$ is the Doppler factor. 
In this paper, we consider a limiting case of
free-fall velocity $\beta=\beta_{\rm ff}=(R_{\rm Sh}/(R+h))^{1/2}$, where $R_{\rm Sh}=2GM/c^2\simeq 2.95\times 10^5 \,{\rm cm}$ is the Schwarzschild radius.

Due to a strong beaming of accretion column radiation towards the NS surface, a large fraction of X-ray photons can be intercepted and reprocessed by the atmosphere of a NS.
The process of X-ray reflection can be affected by the local magnetic field of a NS, but it is poorly investigated up to date.
To model the reflection process, we assume that the intensity of reflected radiation is independent on the direction, i.e. the flux of reflected photons 
\beq 
F_{\rm ref}\propto \cos\theta_{\rm ref},
\eeq 
where $\theta_{\rm ref}$ is the angle between the direction of reflection and local normal to the NS surface.
The flux of the reprocessed radiation is assumed to be equal to the flux locally absorbed by the atmosphere. 

A distant observer detects a flux composed of two components: the direct flux from accretion columns and the flux intercepted and reprocessed by the NS surface (see Fig.\,\ref{pic:scheme_rot}). 
Both components of X-ray flux are subject of gravitational bending in a space curved by a NS.
We assume that the space around a neutron star is described by the Schwarzschild metric. This is a good approximation for neutron stars in XRPs, where the rotation periods of a neutron star are usually longer than $0.1\,{\rm s}$.
Photon trajectories in the Schwarzschild metric are flat.
We describe photon trajectory in polar coordinates $r\geq 0$ and $\varphi\in[0,2\pi]$.
The trajectory of a photon is described by the second order differential equation \citep{1973grav.book.....M}:
\beq 
\frac{\d^2 u}{\d\varphi^2} = 3u^2 - u,
\eeq
where $u=0.5 R_{\rm Sh}/r$ is the compactness of a central object.

\subsection{Mass accretion rate variability}
\label{sec:variability}

The mass accretion rate at the magnetospheric radius experiences stochastic fluctuations. 
The fluctuations of the mass accretion rate in X-ray binaries are caused by clumpiness of the stellar wind \citep{2007A&A...476..335W} or viscosity fluctuations in accretion disc \citep{1997MNRAS.292..679L}.
In the case of accretion from the disc, each radial coordinate produces fluctuation of a certain frequency.
The frequency of locally produced fluctuations is expected to be close to the local Keplerian frequency \citep{2004MNRAS.348..111K}
\beq
f_{\rm K}=\frac{1}{2\pi}\left(\frac{r^3}{GM}\right)^{1/2}.
\eeq
Viscosity fluctuations cause fluctuations of the mass accretion rate that propagate inwards and outwards \citep{2018MNRAS.474.2259M} modulating fluctuations at other radial coordinates.
The fluctuations of the highest Fourier frequency are generated close to the inner radius of accretion disc 
\beq
R_{\rm m}=1.8\times 10^8\,\Lambda\,B_{12}^{4/7}\dot{M}_{17}^{-2/7}m^{-1/7}R_6^{12/7}\,\,{\rm cm},
\eeq
where $B_{12}$ is the magnetic field strength at the magnetic poles of a NS in units of $10^{12}\,{\rm G}$, $\dot{M}_{17}$ is the mass accretion rate in units of $10^{17}\,{\rm g\,s^{-1}}$, $m$ is a mass of a NS in units of solar masses, $R_6$ is the NS radius in units of $10^6\,{\rm cm}$ and $\Lambda$ is a dimensionless coefficient, which is typically taken to be $\Lambda=1$ for the case of accretion from the wind and $\Lambda=0.5$ for the case of accretion from the disc \citep{1978ApJ...223L..83G,1979ApJ...232..259G}.
Thus, the highest Fourier frequency generated in the accretion disc is expected to be 
\beq 
f_{\rm K,max}\sim 0.5\,\Lambda^{-3/2}B_{12}^{-6/7}L_{37}^{3/7}m^{2/7}R_6^{-15/7}\,{\rm Hz}.
\eeq 
According to the propagating fluctuations model applied to the case of XRPs \citep{2019MNRAS.486.4061M}  and observational results \citep{2009A&A...507.1211R,2022MNRAS.515..571M}, the power density spectra of stochastic fluctuations can be described by a broken power law 
function with a break 
at frequency $f\sim f_{\rm K,max}$.
In our simulation, we use this model and assume that the mass accretion rate at the NS surface replicates the fluctuations at the inner disc radius.

\subsection{Neutron star rotation}

NS rotation is described by two parameters: NS inclination $i$ (the angle between the line of sight and the rotational axis) and the magnetic obliquity $\theta_{\rm B}$ (the angle between the magnetic and rotational axis, see Fig.\,\ref{pic:scheme_rot}).
The variations of the angle between the line of sight and magnetic field axis $\psi$ is described by 
\beq \label{eq:psi}
\cos\psi = 
\sin i\sin\theta_B \cos\varphi_{\rm p}+\cos i \cos\theta_B,
\eeq 
where $\varphi_{\rm p}\in[0,2\pi)$ is the phase angle of pulsations.
Typically, the angles $i$ and $\theta_B$ are unknown, but the recent detection of X-ray polarisation variability in XRPs provides an opportunity to probe the rotation geometry (see, e.g., \citealt{2022NatAs...6.1433D,2022ApJ...941L..14T,2023A&A...675A..48T,2023MNRAS.524.2004M,2023A&A...678A.119S,2023A&A...675A..29M}).
Results obtained with the IXPE mission \citep{2022JATIS...8b6002W} demonstrate that the rotation axis of NS in XRPs tends to be orthogonal to the orbital plane of a binary system and, probably, to accretion disc plane.

\section{Numerical model}
\label{sec:Num}

In our simulations, we fix the mass and radius of a NS at $M=1.4M_\odot$ and $R=12\,{\rm km}$ respectively.
The relation between accretion column height and luminosity is described by approximate equation (\ref{eq:L2H_relation}).
Our numerical calculations are divided into a few steps:
\begin{enumerate}[leftmargin=12pt]
\item\label{step:calc_ang_dist}
Following the algorithms described in \citealt{2023MNRAS.tmp.3130M}, we get the  angular distributions of X-ray energy flux for different heights $H$ of accretion column and, therefore, accretion luminosities.
The considered range of accretion column height is $0.05R<H<R$, and to calciulate a database of flux angular distribution we take $100$ height values within the considered interval.
Calculating angular distribution, we account both for direct and reflected flux of accretion column.
Angular distributions of X-ray energy flux for various heights of accretion column and luminosities are tabulated. 
\item
On the base of the assumed power density spectra (PDS) of the mass accretion rate variability and following the algorithm described by \citealt{1995A&A...300..707T}, we simulate a time series $\dot{M}(t)$ that describes mass accretion rate fluctuations.
\item\label{step:L_and_H}
The simulated time series $\dot{M}(t)$ is transformed into a time series for accretion luminosity $L(t)=GM\dot{M}(t)/R$ and accretion column height $H(t)$ according to (\ref{eq:L2H_relation}).
\item
On the base of time series obtained at step \ref{step:L_and_H} and interpolating between the angular distributions of X-ray energy flux calculated in step \ref{step:calc_ang_dist} we get a set of time series of X-ray flux variability in different directions with respect to the NS magnetic axis.
\item
Using the parameters of NS rotation $i$ and $\theta_B$ in the observer's reference frame, we get a model of a pulse profile $F(t,i,\theta_B)$ that fluctuates due to the variable total luminosity and variable geometry of the emission regions (i.e., accretion column and illuminated part of a NS surface). 
\item 
On the base of calculated light curves $F(t,i,\theta_B)$, we get the phase-resolved average flux:
\beq 
F_P(t,i,\theta_B) = \frac{1}{N}\sum\limits_{j=0}^{N-1} F(t+jP,i,\theta_B),
\eeq 
and the phase-resolved rms:
\beq \label{eq:rms}
{\rm rms}_P(t,i,\theta_B) = \frac{1}{N}\sum\limits_{j=0}^{N-1} F^2(t+jP,i,\theta_B)\,-\,F^2_P(t,i,\theta_B),
\eeq 
where $N\gg 1$ is a number of spin period that covers the simulated time series. 
For this paper, we have simulated time series of $10^5 P$ long (i.e., $N=10^5$).
\end{enumerate}

\begin{figure}
\centering 
\includegraphics[width=8.7cm]{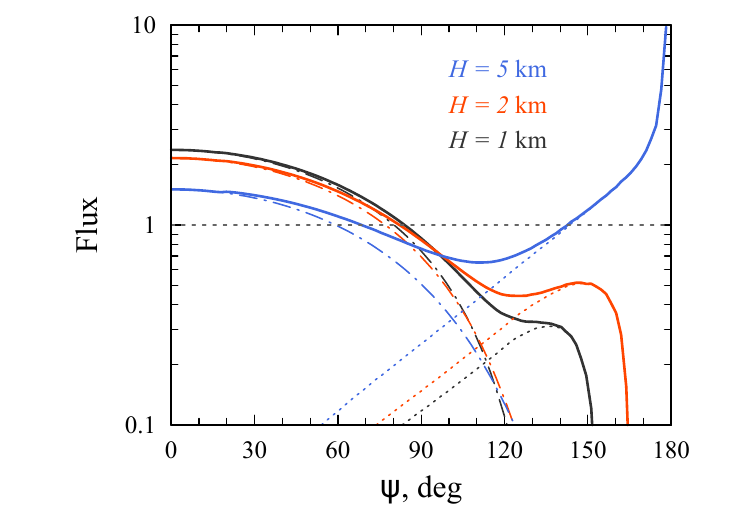} 
\caption{
The angular distributions of X-ray flux {(in units of isotropic flux)} in super-critical XRP with a single accretion column.
The flux is composed of the direct component (dotted lines) and the component reflected from the NS surface (dashed-dotted lines).
{Black, red and blue lines demonstrate the distributions for accretion column height $H=1,\,2$ and $5$ km respectively.
One can see that relatively low accretion columns can be eclipsed by a NS, while high accretion column is gravitationally lensed for observers located on the opposite side from a NS.
Parameters: $M=1.4M_\odot$, $R=12\,{\rm km}$, uniform distribution of emitted flux over accretion column height.}
}
\label{pic:flux_dist}
\end{figure}

\begin{figure*}
\centering 
\includegraphics[width=18.cm]{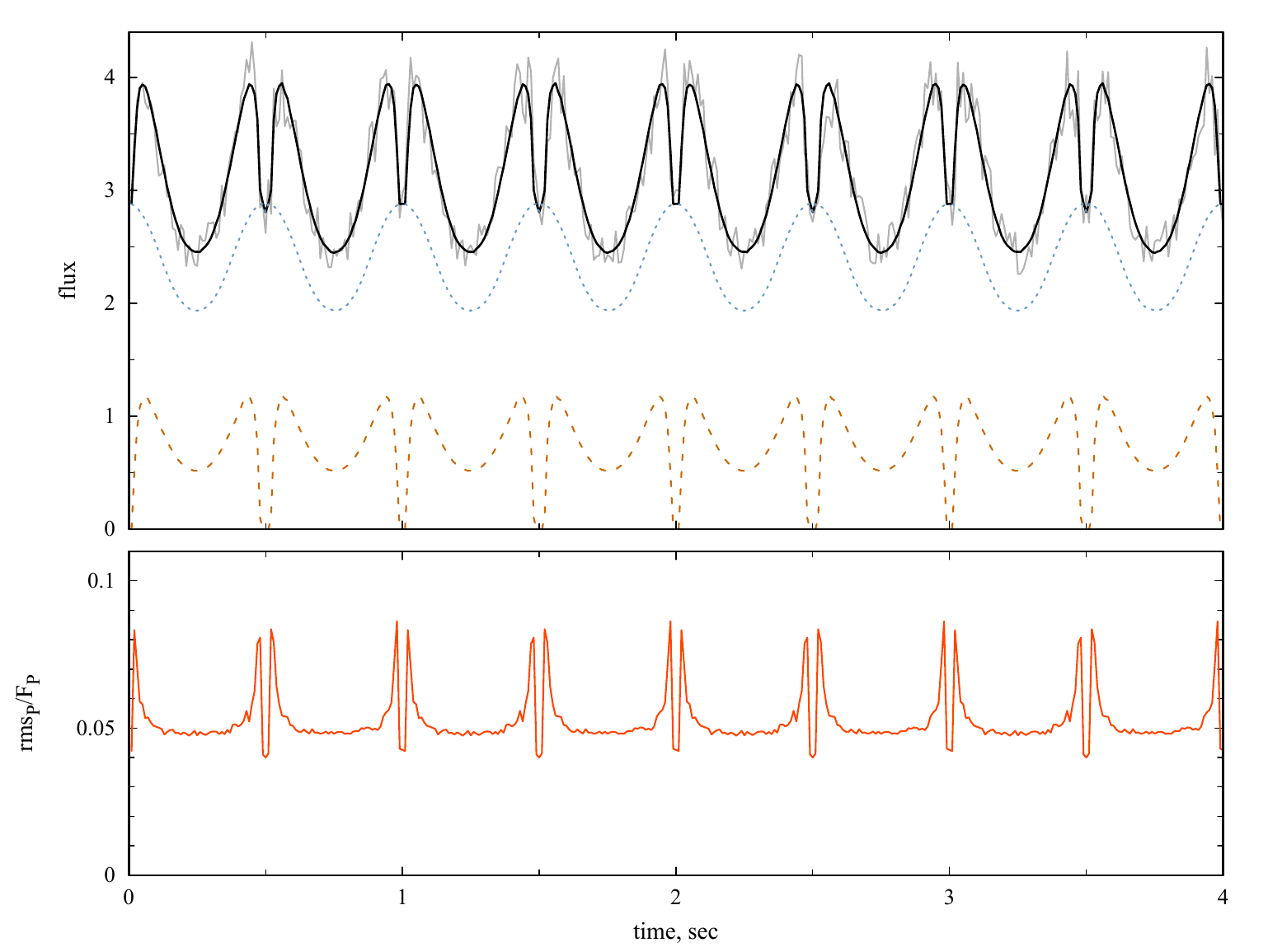} 
\caption{
The simulated pulse profile for a super-critical XRP.
The upper panel demonstrates the fluctuating pulse profile (grey curve) and the averaged pulse profile (black solid line).
The component of the averaged pulse profile due to the direct and reflected emission of accretion column are represented by red dashed and blue dotted curves respectively.
The lower panel demonstrates variations of relative rms over the pulse period. 
The pulse experience the strongest variations just before and right after eclipse of accretion column by a NS. 
Parameters: 
$P=1\,{\rm s}$,
$L=7\times 10^{37}\,\ergs$,
$i=\theta_B=\pi/2$,
${\rm rms}_{\rm ini}=0.06$.
}
\label{pic:pp}
\end{figure*}

\begin{figure}
\centering 
\includegraphics[width=8.cm]{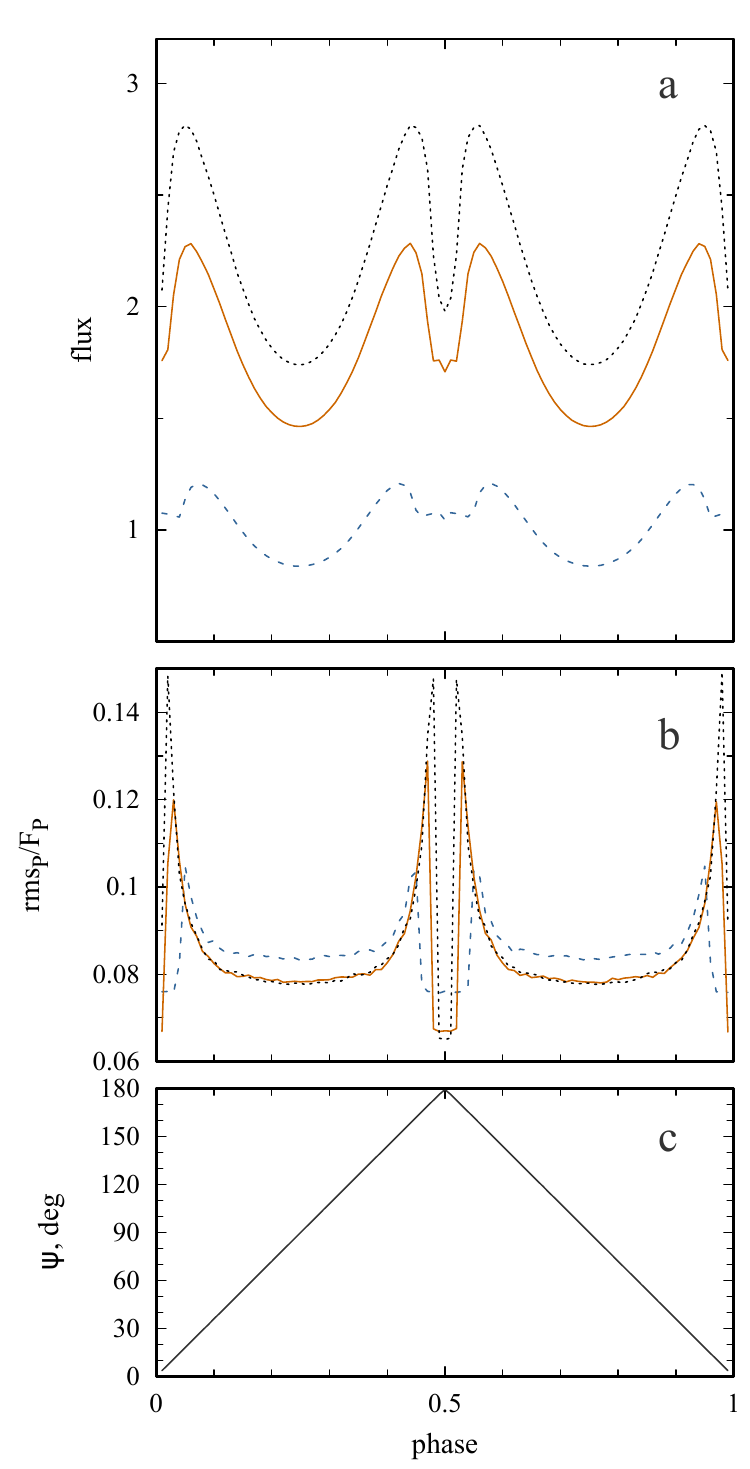} 
\caption{
Variations of flux (a), relative phase-resolved rms (b) and the angle between the line of sight and magnetic field axis $\psi$ (c) over the pulse period.  
Different curves correspond to different accretion column height:
 $0.2R$ (dashed blue), $0.28R$ (solid red) and $0.32R$ (dotted black).
Parameters: 
$M=1.4M_\odot$, $R=12\,{\rm km}$,
$i=\theta_B=\pi/2$,
${\rm rms}_{\rm i}=0.1$.
}
\label{pic:pp_2}
\end{figure}

\begin{figure}
\centering 
\includegraphics[width=8.cm]{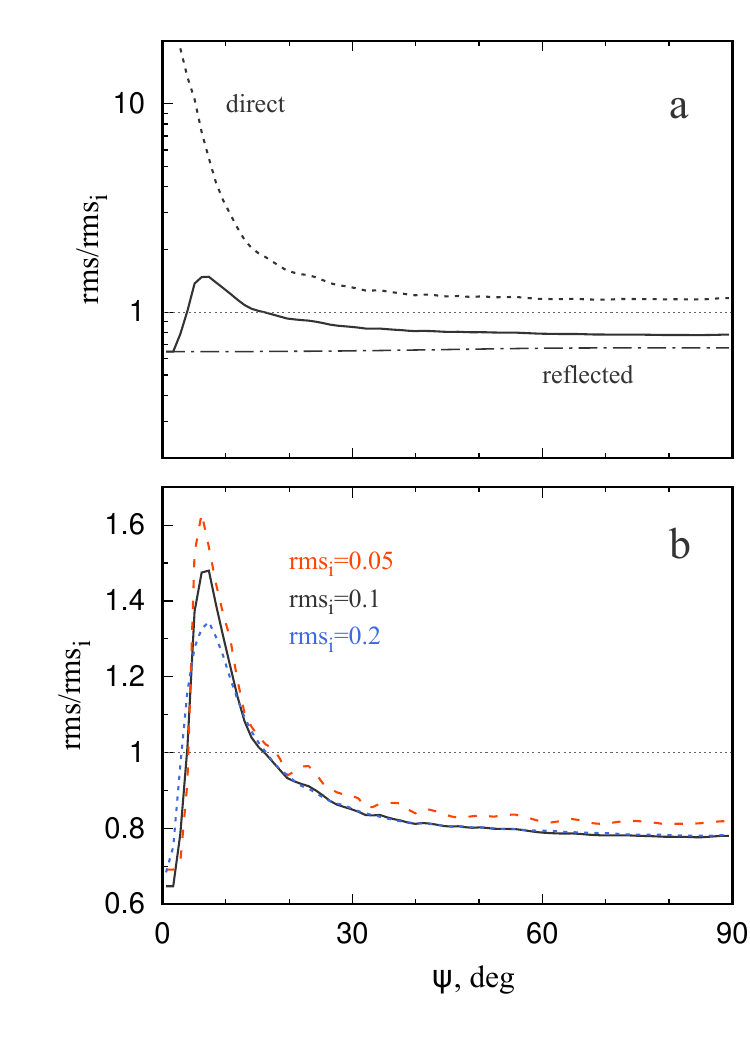} 
\caption{
The relative rms in units of relative rms of the mass accretion rate at a given direction with respect to the NS magnetic axis.
(a) Black line accounts both for direct and reflected flux, while dotted (dashed-dotted) line accounts for direct (reflected) flux only.
Direct flux tend to increase rms, while the reflected componets reduces rms of X-ray energy flux. 
(b) Different lines correspond to different relative rms of the mass accretion rate: $0.05$ (red dashed), $0.1$ (black solid) and $0.2$ (blue dotted).
Parameters: $H=0.32R$, 
$M=1.4M_\odot$, $R=12\,{\rm km}$
}
\label{pic:sc_rms}
\end{figure}

\section{Numerical results}
\label{sec:NumRes}

Both direct and reflected photons experience gravitational bending. 
Due to the gravitational bending, accretion column can illuminate more than a half of a NS surface. 
The examples of X-ray flux angular distributions at the infinity are given in Fig.\,\ref{pic:flux_dist}.
The direct flux from accretion column can be eclipsed by a NS for observers in certain directions (in the case of relatively low accretion columns, see red and blue lines in Fig.\,\ref{pic:flux_dist}) or strongly amplified due to the gravitational lensing (in the case of high accretion column, see black line in Fig.\,\ref{pic:flux_dist}).
The calculations of X-ray flux distributions performed for this paper are in agreement with earlier calculations reported in \citealt{2018MNRAS.474.5425M} and \citealt{2023MNRAS.tmp.3130M} {(see also useful approximations proposed for gravitational bending by \citealt{2002ApJ...566L..85B,2020A&A...640A..24P})}.
X-ray energy flux in a given direction with respect to the magnetic field axis is dependent both on total luminosity and specific geometry of the emitting regions (accretion column and illuminated part of a NS surface).
The direct component of the X-ray flux is more sensitive to the variations of accretion column height than the reflected component (compare dotted and dashed-dotted lines in Fig.\,\ref{pic:flux_dist}). 

The pulse profiles are determined by the angular distribution of X-ray flux in the NS reference frame, and NS rotation in the reference frame of a distant observer, which is given by the inclination $i$ and magnetic obliquity $\theta_B$ (see Fig.\,\ref{pic:scheme_rot}).
To illustrate the relation between specific orientations of a NS given by the angle $\psi$ (see Fig.\,\ref{pic:scheme_rot}) and phase-resolved rms, we demonstrate the pulse profiles for the case of $i=\theta_B=\pi/2$ (see Fig.,\,\ref{pic:pp}, \ref{pic:pp_2}), {when the angle $\psi$ runs through all possible values during one pulsation period.}

Because of the variability in geometry of the emitting region, the rms of X-ray energy flux can be different from the rms of the mass accretion rate. 
The flux variability can be both stronger and weaker than the variability of the mass accretion rate depending on the observer's line of sight (see Fig.\,\ref{pic:sc_rms}).
Fluctuation of reflected flux is typically weaker than the corresponding fluctuations of the mass accretion rate (compare dashed-dotted and solid lines in Fig.\,\ref{pic:sc_rms}a). 
The fluctuation of the direct flux tends to show larger relative rms in comparison to the rms of mass accretion rate fluctuations (compare dotted and solid lines in Fig.\,\ref{pic:sc_rms}a).
Amplification of fluctuations by the direct component are dependent on rms of the mass accretion rate (see Fig.\,\ref{pic:sc_rms}b).

\section{Observational verification}
\label{sec:Obs}

The proposed model was applied to the {\it RXTE}/PCA \citep{1996SPIE.2808...59J} observations of the bright transient XRP V~0332+53 obtained near the peak of the 2004 giant outburst. 
The source is known to enter the supercritical regime of accretion in the bright state \citep{2017MNRAS.466.2143D}, demonstrating the appearance of sharp features in the pulse profile possibly associated with the eclipses of accretion column by NS \citep{2018MNRAS.474.5425M}.

Motivated by the similarities between the model pulse profiles shown in Fig.~\ref{pic:pp} and the observed ones in V~0332+53 we extracted a number of 3-20 keV lightcurves for the source covering a broad range of luminosities from $3\times10^{37}$ to $3\times10^{38}\,\ergs$ (ObsIDs: 90427-01-04-01, 01-03-09, 01-02-03, 01-01-02). The bolometric fluxes (3-100~keV) were adopted from \cite{2015MNRAS.448.2175L} and rescaled to the {\it Gaia} distance of 5.6 kpc \citep{2021AJ....161..147B}. After the barycentric correction, each lightcurve was then folded with a spin period obtained for a given observation. 

\begin{figure}
\centering 
\includegraphics[width=8.5cm]{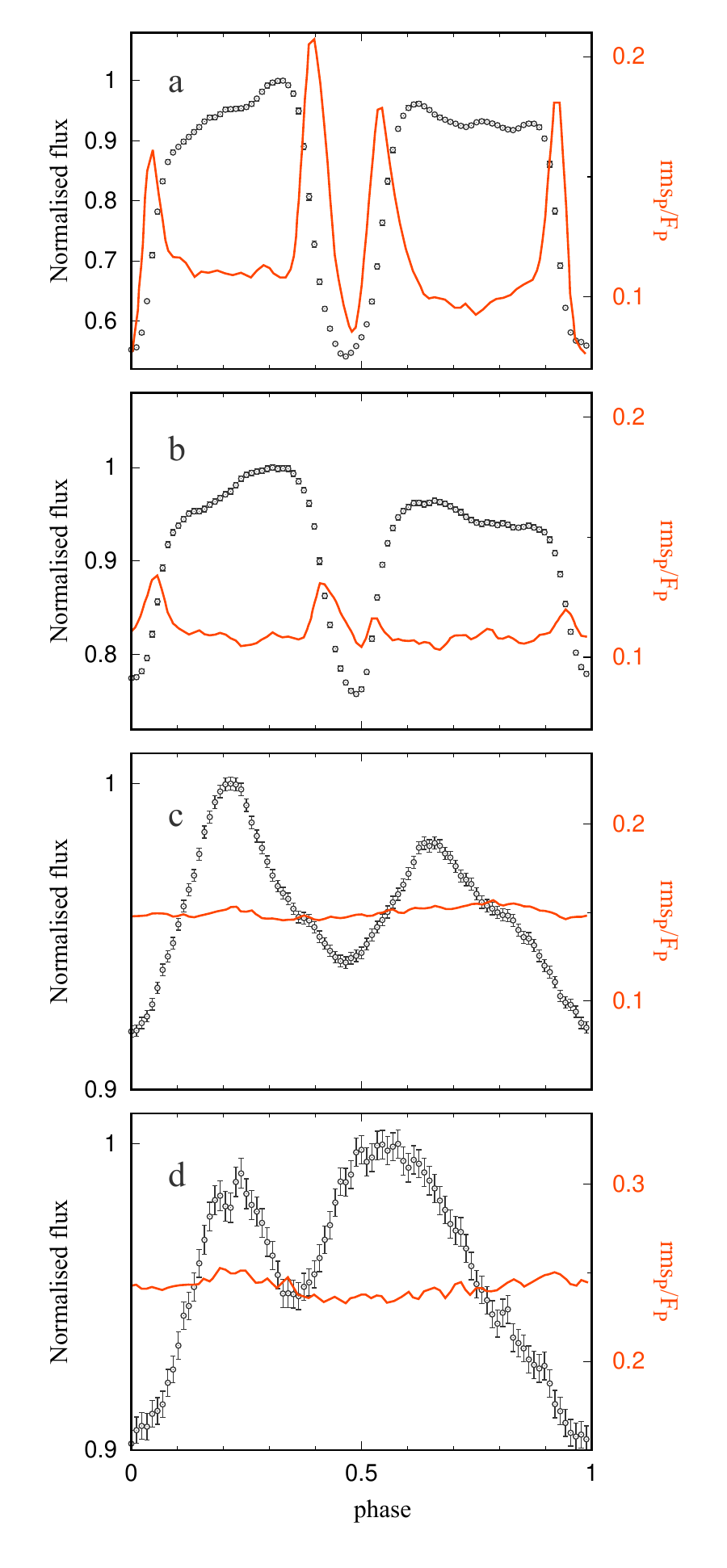} 
\caption{
The pulse profile (depicted by the black circles) observed during a super-critical accretion episode in the XRP V0332+53, along with the normalized phase-resolved root mean square (RMS) shown by the red solid line. Different panels correspond to various accretion luminosities: (a) $3\times 10^{38}\,\ergs$, (b) $2.3\times 10^{38}\,\ergs$, (c) $1.3\times 10^{38}\,\ergs$, and (d) $3.3\times 10^{37}\,\ergs$.
It is evident that the RMS exhibits weak variability with the phase of pulsations at relatively low accretion luminosity (as seen in panels c and d). At high luminosity, the phase-resolved RMS experiences jumps at phases around the troughs in the X-ray pulse profile (as observed in panels a and b). The relative strength of these jumps increases with luminosity, consistent with expectations from the theoretical model (see Fig.\,\ref{pic:pp_2}b).
}
\label{pic:v0332}
\end{figure}

In Fig.~\ref{pic:v0332} we show obtained pulse profiles along with the pulse-phase resolved RMS for four observations at different  X-ray luminosities. 
One can see that the RMS exhibits weak variability with the phase of pulsations at relatively low accretion luminosity (see Fig.~\ref{pic:v0332}c and \ref{pic:v0332}d). 
At a high luminosity state, the RMS shows strong phase dependence experiencing jumps just before and right after the troughs in the X-ray pulse profile (see Fig.\,\ref{pic:v0332}a and \ref{pic:v0332}b). 
The relative strength of these jumps increases with luminosity, consistent with expectations from our theoretical model (see Fig.\,\ref{pic:pp_2}b).



\section{Summary and discussion}
\label{sec:Summary}

We have developed a comprehensive model for pulse profile formation in super-critical XRPs, accounting for the aperiodic variability of the mass accretion rate onto the NS surface. 
In our model, the flux observed by a distant observer is composed of both the direct flux from the accretion column and the flux reflected by the NS's atmosphere (see Fig.\,\ref{pic:scheme}). 
The accretion column geometry, specifically its height in a super-critical state, is considered to be dependent on the mass accretion rate \citep{1976MNRAS.175..395B,2015MNRAS.454.2539M}. 
The height's dependence on the mass accretion rate is calculated following the model proposed in \citealt{2015MNRAS.454.2539M}, but qualitative results and theoretical predictions are weakly dependent on specific features of underling theoretical model. 
Angular distribution of X-ray flux and pulse profiles were computed, incorporating gravitational light bending based on the Schwarzschild metric, a suitable approximation for XRPs due to their relatively slow rotation (typical spin periods $>1\,{\rm s}$).

We simulated pulse profiles that fluctuate due to the variability of the mass accretion rate and the corresponding variability of the accretion column height (see the upper panel in Fig.\,\ref{pic:pp}). 
Utilizing these simulated fluctuating pulse profiles, we calculated the phase-resolved root mean square (rms) of the X-ray energy flux (see eq.\,\ref{pic:pp}). 
Our results demonstrate that accounting for the fluctuating geometry of the emission regions (accretion columns and illuminated part of the NS surface) significantly influences the variability of X-ray energy flux, either amplifying or reducing it with respect to the variability of the mass accretion rate. 
Hence, variability of X-ray energy flux in super-critical XRPs deviates from the variability of the mass accretion rate, even in the reference frame co-rotating with the NS. 
The reduction or amplification of variability depends on the predominant source of X-ray photons. 
Notably, gravitational bending of the direct component can significantly amplify fluctuations, while the flux reflected from the NS surface exhibits reduced variability.

Distinct patterns in phase-resolved rms emerge when the NS eclipses the accretion column at certain phases of pulsations, as depicted in Fig.\,\ref{pic:pp} and \ref{pic:pp_2}: the pulse experiences strong variations just before and right after the eclipse of accretion column by a NS.
Such an amplification of the rms before and after the eclipse of the accretion column are related to gravitational lensing of the column and they appear regardless of exactly how the brightness is distributed over the height of the accretion column. 
In this sense, the prediction of these patterns does not depend on a particular model of the accretion column.
Identifying these patterns in the phase-resolved rms serves as confirmation of accretion column eclipses in a reference frame of a distant observer. 

We have analyzed archival data obtained during the outburst of bright Be X-ray transient V~0332+53 and detected the specific patterns pointing to accretion column eclipses in phase-resolved rms (see Section\,\ref{sec:Obs}) at accretion luminosities exceeding $2\times 10^{38}\,\ergs$ (see Fig.\,\ref{pic:v0332}ab). 
Detected increases in phase-resolved rms align with phases where accretion column eclipses were predicted earlier \citep{2018MNRAS.474.5425M}.
Thus, we have on our hands a strong confirmation that in this source we observe an eclipse of an extended accretion column by a NS.

A confirmed eclipse of the accretion column imposes constraints on (i) the NS's rotation, and (ii) relation between NS mass, radius and accretion column height.
Indeed, appearance of the eclipses requires sufficiently small angle $\psi$ between the line of sight and the magnetic field axis during eclipse phases (see Eq.\,\ref{eq:psi}).
It becomes possible if the inclination of a NS and magnetic obliquity are close to each other (see Fig.\,\ref{pic:scheme_rot}): 
\beq
i\approx \theta_B.
\eeq
This conclusion for a NS in XRP V~0332+53 can be verified by future observations of phase-resolved polarimetry, which becomes possible with NASA's Imaging X-ray Polarimetry Explorer \citep{2022NatAs...6.1433D}.
The X-ray energy spectra during eclipses, mainly composed of photons reflected by the NS's atmosphere, present an opportunity for detailed phase-resolved spectroscopy \citep{2015MNRAS.448.2175L} to validate models of X-ray reflection from magnetized atmospheres.

Because of gravitational light bending in a space curved by a NS, accretion columns can be eclipsed in the case of sufficiently small height and/or small compactness of a star.
Therefore, detecting eclipses in the accretion column imposes constraints on the mass-radius-accretion column height relation in XRPs (refer to Fig.\,17 in \citealt{2018MNRAS.474.5425M}):
\beq
\label{eq:R2M_limit}
R \gtrsim \frac{M}{1.4M_\odot}\left[ 8.7 + 9.8\left(\frac{H}{R}\right) \right]\,{\rm km}.
\eeq
Accurately limiting the NS radius demands precise estimation of the accretion column height, a challenging task with current theoretical models. However, expression (\ref{eq:R2M_limit}) serves as a stringent constraint on the accretion column height and says that the eclipses can only be observed under the condition:
\beq
\frac{H}{R} \lesssim \left(\frac{R}{10\,{\rm km}}\right)\left(\frac{M}{1.4 M_\odot}\right)^{-1} - 0.9.
\eeq
For realistic equations of state \citep{2001ApJ...550..426L} and assuming $R\lesssim 12\,{\rm km}$ and $M>M_\odot$, the detection of eclipses sets a limit on the column height in V~0332+53 for $H/R \lesssim 0.5$ at accretion luminosity as high as $3\times 10^{38}\,\ergs$.
This result is particularly important for the verification of accretion column models, that provide us with the different height-luminosity-magnetic field strength dependence in XRPs.

\section*{Acknowledgements}

We thank UKRI Stephen Hawking fellowship (AAM) and the V\"ais\"al\"a Foundation (SST).

\section*{Data availability}

The calculations presented in this paper were performed using a private code developed and owned by the corresponding author. All the data appearing in the figures are available upon request. 


\begin{thebibliography}{}
\makeatletter
\relax
\def\mn@urlcharsother{\let\do\@makeother \do\$\do\&\do\#\do\^\do\_\do\%\do\~}
\def\mn@doi{\begingroup\mn@urlcharsother \@ifnextchar [ {\mn@doi@}
  {\mn@doi@[]}}
\def\mn@doi@[#1]#2{\def\@tempa{#1}\ifx\@tempa\@empty \href
  {http://dx.doi.org/#2} {doi:#2}\else \href {http://dx.doi.org/#2} {#1}\fi
  \endgroup}
\def\mn@eprint#1#2{\mn@eprint@#1:#2::\@nil}
\def\mn@eprint@arXiv#1{\href {http://arxiv.org/abs/#1} {{\tt arXiv:#1}}}
\def\mn@eprint@dblp#1{\href {http://dblp.uni-trier.de/rec/bibtex/#1.xml}
  {dblp:#1}}
\def\mn@eprint@#1:#2:#3:#4\@nil{\def\@tempa {#1}\def\@tempb {#2}\def\@tempc
  {#3}\ifx \@tempc \@empty \let \@tempc \@tempb \let \@tempb \@tempa \fi \ifx
  \@tempb \@empty \def\@tempb {arXiv}\fi \@ifundefined
  {mn@eprint@\@tempb}{\@tempb:\@tempc}{\expandafter \expandafter \csname
  mn@eprint@\@tempb\endcsname \expandafter{\@tempc}}}

\bibitem[\protect\citeauthoryear{{Abolmasov} \& {Lipunova}}{{Abolmasov} \&
  {Lipunova}}{2023}]{2023MNRAS.524.4148A}
{Abolmasov} P.,  {Lipunova} G.,  2023, \mn@doi [\mnras]
  {10.1093/mnras/stad1951}, \href
  {https://ui.adsabs.harvard.edu/abs/2023MNRAS.524.4148A} {524, 4148}

\bibitem[\protect\citeauthoryear{{Asthana}, {Mushtukov}, {Dobrynina}  \&
  {Ognev}}{{Asthana} et~al.}{2023}]{2023MNRAS.522.3405A}
{Asthana} A.,  {Mushtukov} A.~A.,  {Dobrynina} A.~A.,   {Ognev} I.~S.,  2023,
  \mn@doi [\mnras] {10.1093/mnras/stad1118}, \href
  {https://ui.adsabs.harvard.edu/abs/2023MNRAS.522.3405A} {522, 3405}

\bibitem[\protect\citeauthoryear{{Bailer-Jones}, {Rybizki}, {Fouesneau},
  {Demleitner}  \& {Andrae}}{{Bailer-Jones} et~al.}{2021}]{2021AJ....161..147B}
{Bailer-Jones} C.~A.~L.,  {Rybizki} J.,  {Fouesneau} M.,  {Demleitner} M.,
  {Andrae} R.,  2021, \mn@doi [\aj] {10.3847/1538-3881/abd806}, \href
  {https://ui.adsabs.harvard.edu/abs/2021AJ....161..147B} {161, 147}

\bibitem[\protect\citeauthoryear{{Balbus} \& {Hawley}}{{Balbus} \&
  {Hawley}}{1991}]{1991ApJ...376..214B}
{Balbus} S.~A.,  {Hawley} J.~F.,  1991, \mn@doi [\apj] {10.1086/170270}, \href
  {https://ui.adsabs.harvard.edu/abs/1991ApJ...376..214B} {376, 214}

\bibitem[\protect\citeauthoryear{{Basko} \& {Sunyaev}}{{Basko} \&
  {Sunyaev}}{1976}]{1976MNRAS.175..395B}
{Basko} M.~M.,  {Sunyaev} R.~A.,  1976, \mn@doi [\mnras]
  {10.1093/mnras/175.2.395}, \href
  {https://ui.adsabs.harvard.edu/abs/1976MNRAS.175..395B} {175, 395}

\bibitem[\protect\citeauthoryear{{Beloborodov}}{{Beloborodov}}{2002}]{2002ApJ...566L..85B}
{Beloborodov} A.~M.,  2002, \mn@doi [\apjl] {10.1086/339511}, \href
  {https://ui.adsabs.harvard.edu/abs/2002ApJ...566L..85B} {566, L85}

\bibitem[\protect\citeauthoryear{{Brandenburg}, {Nordlund}, {Stein}  \&
  {Torkelsson}}{{Brandenburg} et~al.}{1995}]{1995ApJ...446..741B}
{Brandenburg} A.,  {Nordlund} A.,  {Stein} R.~F.,   {Torkelsson} U.,  1995,
  \mn@doi [\apj] {10.1086/175831}, \href
  {https://ui.adsabs.harvard.edu/abs/1995ApJ...446..741B} {446, 741}

\bibitem[\protect\citeauthoryear{{Doroshenko}, {Tsygankov}, {Mushtukov},
  {Lutovinov}, {Santangelo}, {Suleimanov}  \& {Poutanen}}{{Doroshenko}
  et~al.}{2017}]{2017MNRAS.466.2143D}
{Doroshenko} V.,  {Tsygankov} S.~S.,  {Mushtukov} A.~A.,  {Lutovinov} A.~A.,
  {Santangelo} A.,  {Suleimanov} V.~F.,   {Poutanen} J.,  2017, \mn@doi
  [\mnras] {10.1093/mnras/stw3236}, \href
  {https://ui.adsabs.harvard.edu/abs/2017MNRAS.466.2143D} {466, 2143}

\bibitem[\protect\citeauthoryear{{Doroshenko} et~al.,}{{Doroshenko}
  et~al.}{2022}]{2022NatAs...6.1433D}
{Doroshenko} V.,  et~al., 2022, \mn@doi [Nature Astronomy]
  {10.1038/s41550-022-01799-5}, \href
  {https://ui.adsabs.harvard.edu/abs/2022NatAs...6.1433D} {6, 1433}

\bibitem[\protect\citeauthoryear{{Ghosh} \& {Lamb}}{{Ghosh} \&
  {Lamb}}{1978}]{1978ApJ...223L..83G}
{Ghosh} P.,  {Lamb} F.~K.,  1978, \mn@doi [\apjl] {10.1086/182734}, \href
  {https://ui.adsabs.harvard.edu/abs/1978ApJ...223L..83G} {223, L83}

\bibitem[\protect\citeauthoryear{{Ghosh} \& {Lamb}}{{Ghosh} \&
  {Lamb}}{1979}]{1979ApJ...232..259G}
{Ghosh} P.,  {Lamb} F.~K.,  1979, \mn@doi [\apj] {10.1086/157285}, \href
  {https://ui.adsabs.harvard.edu/abs/1979ApJ...232..259G} {232, 259}

\bibitem[\protect\citeauthoryear{{Gnedin} \& {Sunyaev}}{{Gnedin} \&
  {Sunyaev}}{1973}]{1973A&A....25..233G}
{Gnedin} Y.~N.,  {Sunyaev} R.~A.,  1973, \aap, \href
  {https://ui.adsabs.harvard.edu/abs/1973A&A....25..233G} {25, 233}

\bibitem[\protect\citeauthoryear{{Jahoda}, {Swank}, {Giles}, {Stark},
  {Strohmayer}, {Zhang}  \& {Morgan}}{{Jahoda}
  et~al.}{1996}]{1996SPIE.2808...59J}
{Jahoda} K.,  {Swank} J.~H.,  {Giles} A.~B.,  {Stark} M.~J.,  {Strohmayer} T.,
  {Zhang} W.,   {Morgan} E.~H.,  1996, in {Siegmund} O.~H.,  {Gummin} M.~A.,
  eds,  Society of Photo-Optical Instrumentation Engineers (SPIE) Conference
  Series Vol. 2808, EUV, X-Ray, and Gamma-Ray Instrumentation for Astronomy
  VII. pp 59--70, \mn@doi{10.1117/12.256034}

\bibitem[\protect\citeauthoryear{{Kaminker}, {Fedorenko}  \&
  {Tsygan}}{{Kaminker} et~al.}{1976}]{1976SvA....20..436K}
{Kaminker} A.~D.,  {Fedorenko} V.~N.,   {Tsygan} A.~I.,  1976, \sovast, \href
  {https://ui.adsabs.harvard.edu/abs/1976SvA....20..436K} {20, 436}

\bibitem[\protect\citeauthoryear{{King}, {Pringle}, {West}  \& {Livio}}{{King}
  et~al.}{2004}]{2004MNRAS.348..111K}
{King} A.~R.,  {Pringle} J.~E.,  {West} R.~G.,   {Livio} M.,  2004, \mn@doi
  [\mnras] {10.1111/j.1365-2966.2004.07322.x}, \href
  {https://ui.adsabs.harvard.edu/abs/2004MNRAS.348..111K} {348, 111}

\bibitem[\protect\citeauthoryear{{Kraus}}{{Kraus}}{2001}]{2001ApJ...563..289K}
{Kraus} U.,  2001, \mn@doi [\apj] {10.1086/323791}, \href
  {https://ui.adsabs.harvard.edu/abs/2001ApJ...563..289K} {563, 289}

\bibitem[\protect\citeauthoryear{{Lattimer} \& {Prakash}}{{Lattimer} \&
  {Prakash}}{2001}]{2001ApJ...550..426L}
{Lattimer} J.~M.,  {Prakash} M.,  2001, \mn@doi [\apj] {10.1086/319702}, \href
  {https://ui.adsabs.harvard.edu/abs/2001ApJ...550..426L} {550, 426}

\bibitem[\protect\citeauthoryear{{Lutovinov}, {Tsygankov}, {Suleimanov},
  {Mushtukov}, {Doroshenko}, {Nagirner}  \& {Poutanen}}{{Lutovinov}
  et~al.}{2015}]{2015MNRAS.448.2175L}
{Lutovinov} A.~A.,  {Tsygankov} S.~S.,  {Suleimanov} V.~F.,  {Mushtukov} A.~A.,
   {Doroshenko} V.,  {Nagirner} D.~I.,   {Poutanen} J.,  2015, \mn@doi [\mnras]
  {10.1093/mnras/stv125}, \href
  {https://ui.adsabs.harvard.edu/abs/2015MNRAS.448.2175L} {448, 2175}

\bibitem[\protect\citeauthoryear{{Lyubarskii}}{{Lyubarskii}}{1997}]{1997MNRAS.292..679L}
{Lyubarskii} Y.~E.,  1997, \mn@doi [\mnras] {10.1093/mnras/292.3.679}, \href
  {https://ui.adsabs.harvard.edu/abs/1997MNRAS.292..679L} {292, 679}

\bibitem[\protect\citeauthoryear{{Lyubarskii} \& {Syunyaev}}{{Lyubarskii} \&
  {Syunyaev}}{1988}]{1988SvAL...14..390L}
{Lyubarskii} Y.~E.,  {Syunyaev} R.~A.,  1988, Soviet Astronomy Letters, \href
  {https://ui.adsabs.harvard.edu/abs/1988SvAL...14..390L} {14, 390}

\bibitem[\protect\citeauthoryear{{Malacaria} et~al.,}{{Malacaria}
  et~al.}{2023}]{2023A&A...675A..29M}
{Malacaria} C.,  et~al., 2023, \mn@doi [\aap] {10.1051/0004-6361/202346581},
  \href {https://ui.adsabs.harvard.edu/abs/2023A&A...675A..29M} {675, A29}

\bibitem[\protect\citeauthoryear{{Markozov} \& {Mushtukov}}{{Markozov} \&
  {Mushtukov}}{2024}]{2024MNRAS.527.5374M}
{Markozov} I.~D.,  {Mushtukov} A.~A.,  2024, \mn@doi [\mnras]
  {10.1093/mnras/stad3248}, \href
  {https://ui.adsabs.harvard.edu/abs/2024MNRAS.527.5374M} {527, 5374}

\bibitem[\protect\citeauthoryear{{Meszaros}}{{Meszaros}}{1992}]{1992herm.book.....M}
{Meszaros} P.,  1992, {High-energy radiation from magnetized neutron stars}

\bibitem[\protect\citeauthoryear{{Misner}, {Thorne}  \& {Wheeler}}{{Misner}
  et~al.}{1973}]{1973grav.book.....M}
{Misner} C.~W.,  {Thorne} K.~S.,   {Wheeler} J.~A.,  1973, {Gravitation}

\bibitem[\protect\citeauthoryear{{M{\"o}nkk{\"o}nen}, {Tsygankov}, {Mushtukov},
  {Doroshenko}, {Suleimanov}  \& {Poutanen}}{{M{\"o}nkk{\"o}nen}
  et~al.}{2022}]{2022MNRAS.515..571M}
{M{\"o}nkk{\"o}nen} J.,  {Tsygankov} S.~S.,  {Mushtukov} A.~A.,  {Doroshenko}
  V.,  {Suleimanov} V.~F.,   {Poutanen} J.,  2022, \mn@doi [\mnras]
  {10.1093/mnras/stac1828}, \href
  {https://ui.adsabs.harvard.edu/abs/2022MNRAS.515..571M} {515, 571}

\bibitem[\protect\citeauthoryear{{Mushtukov} \& {Tsygankov}}{{Mushtukov} \&
  {Tsygankov}}{2022}]{2022arXiv220414185M}
{Mushtukov} A.,  {Tsygankov} S.,  2022, \mn@doi [arXiv e-prints]
  {10.48550/arXiv.2204.14185}, \href
  {https://ui.adsabs.harvard.edu/abs/2022arXiv220414185M} {p. arXiv:2204.14185}

\bibitem[\protect\citeauthoryear{{Mushtukov}, {Suleimanov}, {Tsygankov}  \&
  {Poutanen}}{{Mushtukov} et~al.}{2015a}]{2015MNRAS.447.1847M}
{Mushtukov} A.~A.,  {Suleimanov} V.~F.,  {Tsygankov} S.~S.,   {Poutanen} J.,
  2015a, \mn@doi [\mnras] {10.1093/mnras/stu2484}, \href
  {https://ui.adsabs.harvard.edu/abs/2015MNRAS.447.1847M} {447, 1847}

\bibitem[\protect\citeauthoryear{{Mushtukov}, {Suleimanov}, {Tsygankov}  \&
  {Poutanen}}{{Mushtukov} et~al.}{2015b}]{2015MNRAS.454.2539M}
{Mushtukov} A.~A.,  {Suleimanov} V.~F.,  {Tsygankov} S.~S.,   {Poutanen} J.,
  2015b, \mn@doi [\mnras] {10.1093/mnras/stv2087}, \href
  {https://ui.adsabs.harvard.edu/abs/2015MNRAS.454.2539M} {454, 2539}

\bibitem[\protect\citeauthoryear{{Mushtukov}, {Ingram}  \& {van der
  Klis}}{{Mushtukov} et~al.}{2018a}]{2018MNRAS.474.2259M}
{Mushtukov} A.~A.,  {Ingram} A.,   {van der Klis} M.,  2018a, \mn@doi [\mnras]
  {10.1093/mnras/stx2872}, \href
  {https://ui.adsabs.harvard.edu/abs/2018MNRAS.474.2259M} {474, 2259}

\bibitem[\protect\citeauthoryear{{Mushtukov}, {Verhagen}, {Tsygankov}, {van der
  Klis}, {Lutovinov}  \& {Larchenkova}}{{Mushtukov}
  et~al.}{2018b}]{2018MNRAS.474.5425M}
{Mushtukov} A.~A.,  {Verhagen} P.~A.,  {Tsygankov} S.~S.,  {van der Klis} M.,
  {Lutovinov} A.~A.,   {Larchenkova} T.~I.,  2018b, \mn@doi [\mnras]
  {10.1093/mnras/stx2905}, \href
  {https://ui.adsabs.harvard.edu/abs/2018MNRAS.474.5425M} {474, 5425}

\bibitem[\protect\citeauthoryear{{Mushtukov}, {Tsygankov}, {Suleimanov}  \&
  {Poutanen}}{{Mushtukov} et~al.}{2018c}]{2018MNRAS.476.2867M}
{Mushtukov} A.~A.,  {Tsygankov} S.~S.,  {Suleimanov} V.~F.,   {Poutanen} J.,
  2018c, \mn@doi [\mnras] {10.1093/mnras/sty379}, \href
  {https://ui.adsabs.harvard.edu/abs/2018MNRAS.476.2867M} {476, 2867}

\bibitem[\protect\citeauthoryear{{Mushtukov}, {Ognev}  \&
  {Nagirner}}{{Mushtukov} et~al.}{2019a}]{2019MNRAS.485L.131M}
{Mushtukov} A.~A.,  {Ognev} I.~S.,   {Nagirner} D.~I.,  2019a, \mn@doi [\mnras]
  {10.1093/mnrasl/slz047}, \href
  {https://ui.adsabs.harvard.edu/abs/2019MNRAS.485L.131M} {485, L131}

\bibitem[\protect\citeauthoryear{{Mushtukov}, {Lipunova}, {Ingram},
  {Tsygankov}, {M{\"o}nkk{\"o}nen}  \& {van der Klis}}{{Mushtukov}
  et~al.}{2019b}]{2019MNRAS.486.4061M}
{Mushtukov} A.~A.,  {Lipunova} G.~V.,  {Ingram} A.,  {Tsygankov} S.~S.,
  {M{\"o}nkk{\"o}nen} J.,   {van der Klis} M.,  2019b, \mn@doi [\mnras]
  {10.1093/mnras/stz948}, \href
  {https://ui.adsabs.harvard.edu/abs/2019MNRAS.486.4061M} {486, 4061}

\bibitem[\protect\citeauthoryear{{Mushtukov} et~al.,}{{Mushtukov}
  et~al.}{2023}]{2023MNRAS.524.2004M}
{Mushtukov} A.~A.,  et~al., 2023, \mn@doi [\mnras] {10.1093/mnras/stad1961},
  \href {https://ui.adsabs.harvard.edu/abs/2023MNRAS.524.2004M} {524, 2004}

\bibitem[\protect\citeauthoryear{{Postnov}, {Gornostaev}, {Klochkov},
  {Laplace}, {Lukin}  \& {Shakura}}{{Postnov}
  et~al.}{2015}]{2015MNRAS.452.1601P}
{Postnov} K.~A.,  {Gornostaev} M.~I.,  {Klochkov} D.,  {Laplace} E.,  {Lukin}
  V.~V.,   {Shakura} N.~I.,  2015, \mn@doi [\mnras] {10.1093/mnras/stv1393},
  \href {https://ui.adsabs.harvard.edu/abs/2015MNRAS.452.1601P} {452, 1601}

\bibitem[\protect\citeauthoryear{{Pottschmidt} et~al.,}{{Pottschmidt}
  et~al.}{2005}]{2005ApJ...634L..97P}
{Pottschmidt} K.,  et~al., 2005, \mn@doi [\apjl] {10.1086/498689}, \href
  {https://ui.adsabs.harvard.edu/abs/2005ApJ...634L..97P} {634, L97}

\bibitem[\protect\citeauthoryear{{Poutanen}}{{Poutanen}}{2020}]{2020A&A...640A..24P}
{Poutanen} J.,  2020, \mn@doi [\aap] {10.1051/0004-6361/202037471}, \href
  {https://ui.adsabs.harvard.edu/abs/2020A&A...640A..24P} {640, A24}

\bibitem[\protect\citeauthoryear{{Poutanen}, {Mushtukov}, {Suleimanov},
  {Tsygankov}, {Nagirner}, {Doroshenko}  \& {Lutovinov}}{{Poutanen}
  et~al.}{2013}]{2013ApJ...777..115P}
{Poutanen} J.,  {Mushtukov} A.~A.,  {Suleimanov} V.~F.,  {Tsygankov} S.~S.,
  {Nagirner} D.~I.,  {Doroshenko} V.,   {Lutovinov} A.~A.,  2013, \mn@doi
  [\apj] {10.1088/0004-637X/777/2/115}, \href
  {https://ui.adsabs.harvard.edu/abs/2013ApJ...777..115P} {777, 115}

\bibitem[\protect\citeauthoryear{{Revnivtsev}, {Churazov}, {Postnov}  \&
  {Tsygankov}}{{Revnivtsev} et~al.}{2009}]{2009A&A...507.1211R}
{Revnivtsev} M.,  {Churazov} E.,  {Postnov} K.,   {Tsygankov} S.,  2009,
  \mn@doi [\aap] {10.1051/0004-6361/200912317}, \href
  {https://ui.adsabs.harvard.edu/abs/2009A&A...507.1211R} {507, 1211}

\bibitem[\protect\citeauthoryear{{Riffert} \& {Meszaros}}{{Riffert} \&
  {Meszaros}}{1988}]{1988ApJ...325..207R}
{Riffert} H.,  {Meszaros} P.,  1988, \mn@doi [\apj] {10.1086/165996}, \href
  {https://ui.adsabs.harvard.edu/abs/1988ApJ...325..207R} {325, 207}

\bibitem[\protect\citeauthoryear{{Sheng}, {Zhang}, {Blaes}  \& {Jiang}}{{Sheng}
  et~al.}{2023}]{2023MNRAS.524.2431S}
{Sheng} X.,  {Zhang} L.,  {Blaes} O.,   {Jiang} Y.-F.,  2023, \mn@doi [\mnras]
  {10.1093/mnras/stad2043}, \href
  {https://ui.adsabs.harvard.edu/abs/2023MNRAS.524.2431S} {524, 2431}

\bibitem[\protect\citeauthoryear{{Suleimanov}, {Mushtukov}, {Ognev},
  {Doroshenko}  \& {Werner}}{{Suleimanov} et~al.}{2022}]{2022MNRAS.517.4022S}
{Suleimanov} V.~F.,  {Mushtukov} A.~A.,  {Ognev} I.,  {Doroshenko} V.~A.,
  {Werner} K.,  2022, \mn@doi [\mnras] {10.1093/mnras/stac2935}, \href
  {https://ui.adsabs.harvard.edu/abs/2022MNRAS.517.4022S} {517, 4022}

\bibitem[\protect\citeauthoryear{{Suleimanov} et~al.,}{{Suleimanov}
  et~al.}{2023}]{2023A&A...678A.119S}
{Suleimanov} V.~F.,  et~al., 2023, \mn@doi [\aap]
  {10.1051/0004-6361/202346994}, \href
  {https://ui.adsabs.harvard.edu/abs/2023A&A...678A.119S} {678, A119}

\bibitem[\protect\citeauthoryear{{Timmer} \& {Koenig}}{{Timmer} \&
  {Koenig}}{1995}]{1995A&A...300..707T}
{Timmer} J.,  {Koenig} M.,  1995, \aap, \href
  {https://ui.adsabs.harvard.edu/abs/1995A&A...300..707T} {300, 707}

\bibitem[\protect\citeauthoryear{{Tsygankov}, {Lutovinov}, {Churazov}  \&
  {Sunyaev}}{{Tsygankov} et~al.}{2006}]{2006MNRAS.371...19T}
{Tsygankov} S.~S.,  {Lutovinov} A.~A.,  {Churazov} E.~M.,   {Sunyaev} R.~A.,
  2006, \mn@doi [\mnras] {10.1111/j.1365-2966.2006.10610.x}, \href
  {https://ui.adsabs.harvard.edu/abs/2006MNRAS.371...19T} {371, 19}

\bibitem[\protect\citeauthoryear{{Tsygankov}, {Lutovinov}, {Churazov}  \&
  {Sunyaev}}{{Tsygankov} et~al.}{2007}]{2007AstL...33..368T}
{Tsygankov} S.~S.,  {Lutovinov} A.~A.,  {Churazov} E.~M.,   {Sunyaev} R.~A.,
  2007, \mn@doi [Astronomy Letters] {10.1134/S1063773707060023}, \href
  {https://ui.adsabs.harvard.edu/abs/2007AstL...33..368T} {33, 368}

\bibitem[\protect\citeauthoryear{{Tsygankov} et~al.,}{{Tsygankov}
  et~al.}{2022}]{2022ApJ...941L..14T}
{Tsygankov} S.~S.,  et~al., 2022, \mn@doi [\apjl] {10.3847/2041-8213/aca486},
  \href {https://ui.adsabs.harvard.edu/abs/2022ApJ...941L..14T} {941, L14}

\bibitem[\protect\citeauthoryear{{Tsygankov} et~al.,}{{Tsygankov}
  et~al.}{2023}]{2023A&A...675A..48T}
{Tsygankov} S.~S.,  et~al., 2023, \mn@doi [\aap] {10.1051/0004-6361/202346134},
  \href {https://ui.adsabs.harvard.edu/abs/2023A&A...675A..48T} {675, A48}

\bibitem[\protect\citeauthoryear{{Walter} \& {Zurita Heras}}{{Walter} \&
  {Zurita Heras}}{2007}]{2007A&A...476..335W}
{Walter} R.,  {Zurita Heras} J.,  2007, \mn@doi [\aap]
  {10.1051/0004-6361:20078353}, \href
  {https://ui.adsabs.harvard.edu/abs/2007A&A...476..335W} {476, 335}

\bibitem[\protect\citeauthoryear{{Wang} \& {Frank}}{{Wang} \&
  {Frank}}{1981}]{1981A&A....93..255W}
{Wang} Y.~M.,  {Frank} J.,  1981, \aap, \href
  {https://ui.adsabs.harvard.edu/abs/1981A&A....93..255W} {93, 255}

\bibitem[\protect\citeauthoryear{{Weisskopf} et~al.,}{{Weisskopf}
  et~al.}{2022}]{2022JATIS...8b6002W}
{Weisskopf} M.~C.,  et~al., 2022, \mn@doi [Journal of Astronomical Telescopes,
  Instruments, and Systems] {10.1117/1.JATIS.8.2.026002}, \href
  {https://ui.adsabs.harvard.edu/abs/2022JATIS...8b6002W} {8, 026002}

\bibitem[\protect\citeauthoryear{{Zel'dovich} \& {Shakura}}{{Zel'dovich} \&
  {Shakura}}{1969}]{1969SvA....13..175Z}
{Zel'dovich} Y.~B.,  {Shakura} N.~I.,  1969, \sovast, \href
  {https://ui.adsabs.harvard.edu/abs/1969SvA....13..175Z} {13, 175}

\makeatother
\end{thebibliography}

\bsp 
\label{lastpage}
\end{document}